\documentclass[reprint, aps,  prd, usenatbib,twocolumn, amsmath, amssymb, superscriptaddress, nofootinbib, nobibnotes]{revtex4}

\usepackage{natbib}
\usepackage{graphicx}	
\usepackage{amsmath,amsfonts,bm}
\usepackage{hyperref}
\usepackage[T1]{fontenc}
\usepackage[symbol]{footmisc}

\newcommand{\aap}{A\&A}
 
\newcommand{\aj}{Astrophysical Journal}

\newcommand\mnras{MNRAS}

\newcommand{\hmpc}{$h^{-1}{\rm Mpc}$ }

\begin{document}

\title{Arguments against using \hmpc units in observational cosmology
}

\author{Ariel G. S\'anchez}%
\surname{}
\email[ ]{arielsan@mpe.mpg.de}
\affiliation{ Max-Planck-Institut f\"ur extraterrestrische Physik, Postfach 1312, Giessenbachstr., 85741 Garching, Germany}
\date{\today}%

\begin{abstract}
It is common to express cosmological measurements in units of $h^{-1}{\rm Mpc}$. 
Here, we review some of the complications that originate from this practice. 
A crucial problem caused by these units is related to the normalization of the matter power 
spectrum, which is commonly characterized in terms of the linear-theory rms mass fluctuation 
in spheres of radius $8\,h^{-1}{\rm Mpc}$, $\sigma_8$.
This parameter does not correctly capture the impact of $h$ on the amplitude of density fluctuations. 
We show that the use of $\sigma_8$ has caused critical misconceptions for both the so-called 
$\sigma_8$ tension regarding the consistency between low-redshift probes and cosmic microwave 
background data, and the way in which growth-rate estimates inferred from redshift-space 
distortions are commonly expressed. We propose to abandon the use 
of $h^{-1}{\rm Mpc}$ units in cosmology and to characterize the amplitude of the matter power 
spectrum in terms of $\sigma_{12}$, defined as the mass fluctuation in spheres of radius 
$12\,{\rm Mpc}$, whose value is similar to the standard $\sigma_8$ for $h\sim 0.67$.
\end{abstract}

\maketitle

\section{Introduction}
Most statistics used to analyze the large-scale structure of the Universe 
require the assumption of a fiducial cosmology to relate observable quantities 
such as galaxy angular positions and redshifts to density fluctuations on a 
given physical scale. 
To avoid adopting a specific value of the Hubble parameter, it is common 
to express all scales in units of $h^{-1}{\rm Mpc}$, where $h$ determines 
the present-day value of the Hubble parameter as
$H_0 = 100\,h\,{\rm km}\,{\rm s}^{-1}{\rm Mpc}^{-1}$.  At low redshift, 
where the comoving distance, $\chi(z)$, can be approximated as 
\begin{equation}
\chi(z) \approx \frac{c}{H_0}z, \label{eq:comoving_dist}
\end{equation}
using $h^{-1}{\rm Mpc}$ units effectively yields a distance
independent of the fiducial cosmology. This approach
was applied to the analysis of the first galaxy redshift surveys 
\cite{Davis1983, Shectman1996}, which probed only small volumes.
However, this practice has continued until the analysis of present-day
samples such as the Baryon Oscillation Spectroscopic Survey (BOSS) \cite{Dawson2013}, 
which covers a larger redshift range in which computing $\chi(z)$ 
requires the assumption of a full set of fiducial cosmological parameters.

As cosmological observations are expressed in \hmpc  units, theoretical predictions 
follow the same approach.  These units obscure the
dependence of the matter power spectrum, $P(k)$,  on $h$.  
Moreover, the amplitude of $P(k)$ is often characterized in 
terms of the rms linear perturbation theory variance in spheres of radius 
$R=8\,h^{-1}\,{\rm Mpc}$, commonly denoted as $\sigma_8$.
In this paper,  we discuss the misconceptions 
related with the use of \hmpc units and the normalization
of model predictions in terms of $\sigma_8$, and how they can be 
avoided.

\begin{figure*}
\includegraphics[width=0.92\textwidth]{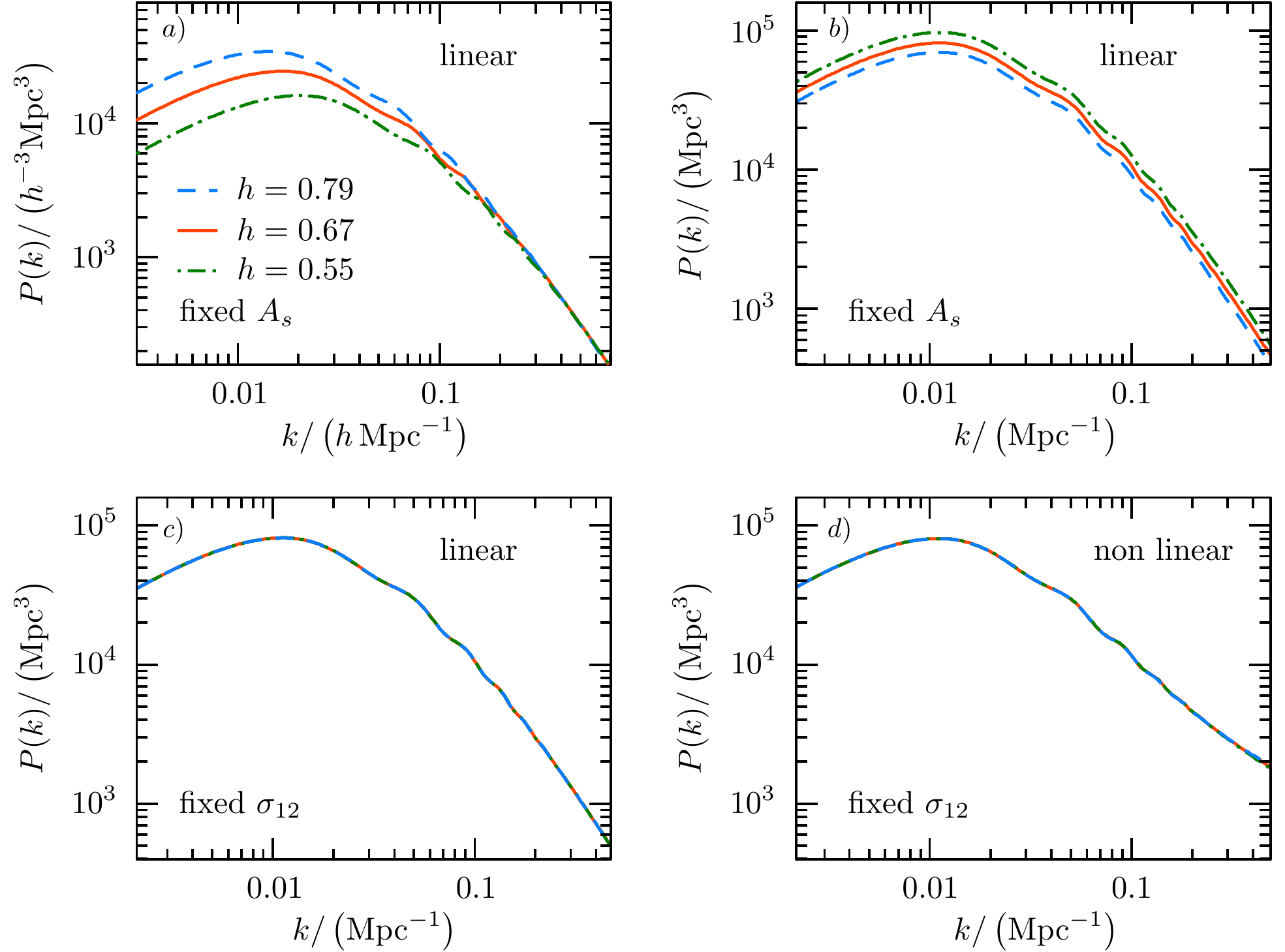}
\caption{\label{fig:normpk} 
Panel $a)$: Linear matter power spectra at $z=0$ of three $\Lambda$CDM models defined
by identical values of $\omega_{\rm b}$, $\omega_{\rm c}$, $\omega_{\nu}$, 
$A_{\rm s}$ and $n_{\rm s}$, and varying $h$, expressed in $h^{-1}{\rm Mpc}$ units. 
Panel $b)$: The same power spectra of panel $a$) shown in ${\rm Mpc}$ units. Panel $c)$: 
The power spectra of the same models of panel $b$) but with their values of $A_{\rm s}$ 
adapted to produce the same value of $\sigma_{12}$.
Panel $d)$: Nonlinear matter power spectra corresponding to the same models 
of panel $c$).}
\end{figure*}

\section{Impact of the fiducial cosmology}
Three-dimensional galaxy clustering 
measurements depend on 
the cosmology used to transform the observed redshifts 
into distances. Any difference between this fiducial cosmology and the 
true one gives rise to the so-called Alcock-Paczynski (AP) distortions \citep{Alcock1979}. 
This geometric effect distorts the 
inferred components parallel and perpendicular to the 
line of sight, $s_{\parallel}$ and $s_{\perp}$, of the 
 separation vector ${\mathbf s}$ between any two 
galaxies as
\citep{Padmanabhan2008,Kazin2012}
\begin{align}
 s_{\parallel} &= q_{\parallel}s'_{\parallel},\label{eq:scaling1}\\
  s_{\perp}     &= q_{\perp}s'_{\perp},\label{eq:scaling2}
\end{align}
where the primes denote the quantities in the fiducial
cosmology, and the scaling factors are given by
\begin{align}
 q_{\parallel} &= \frac{H'(z_{\rm m})}{H(z_{\rm m})},\label{eq:q_para}\\
 q_{\perp} &= \frac{D_{\rm M}(z_{\rm m})}{D'_{\rm M}(z_{\rm m})}, \label{eq:q_perp}
\end{align}
where $H(z)$ is the Hubble parameter,
$D_{\rm M}(z)$ is the comoving angular diameter distance, 
 and  $z_{\rm m}$ is the effective redshift of 
the galaxy sample. 
If the clustering measurements are expressed in $h^{-1}{\rm Mpc}$, the quantities appearing in 
Eqs.~(\ref{eq:q_para}) and (\ref{eq:q_perp}) must also be computed in these units. 

Using $h^{-1}{\rm Mpc}$ units or simply ${\rm Mpc}$ would lead to identical parameter 
constraints, as the factors of $h$ in the model and fiducial cosmologies that enter in $q_{\perp,\parallel}$
would simply cancel out with those in $s_{\perp,\parallel}$ in 
Eqs.~(\ref{eq:scaling1}) and (\ref{eq:scaling2}).
This simply reflects that, when $h$ is correctly taken into account in the 
scaling parameters $q_{\perp,\parallel}$, the constraints derived from clustering 
data are not sensitive to the units in which they are expressed. The fact that clustering measurements
can be expressed in $h^{-1}{\rm Mpc}$ without the explicit assumption of a value of $h$ has no 
impact on the information content of these data, and it does not imply that only quantities 
referred to scales in $h^{-1}{\rm Mpc}$ units can be derived from them.

\begin{figure*}
\includegraphics[width=0.92\textwidth]{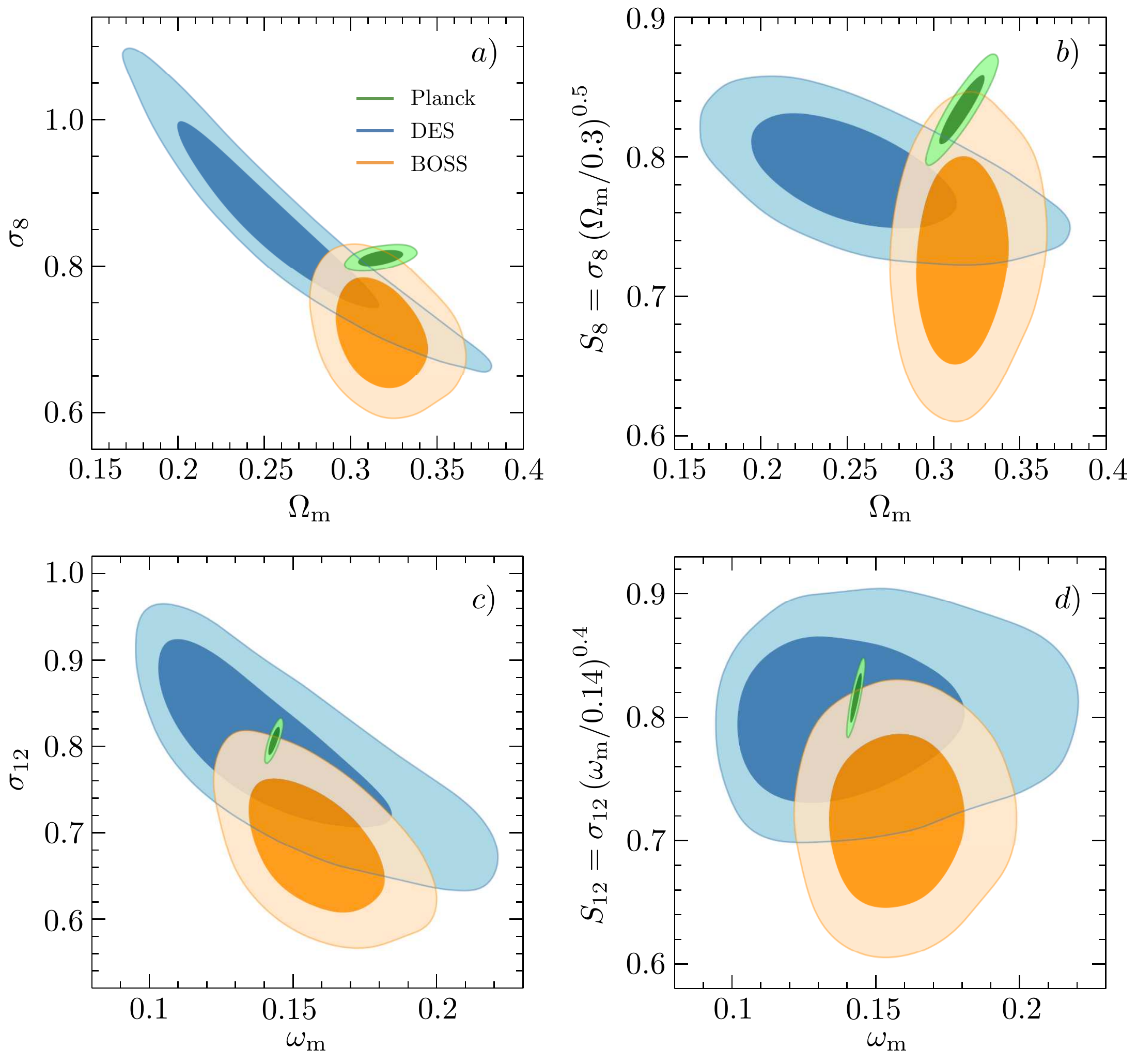}
\caption{\label{fig:consistency} 
Two-dimensional 68\% and 95\% constraints recovered 
from Planck (green), the $3\times 2$pt analysis of DES (blue), 
and BOSS (orange) under the assumption of a $\Lambda$CDM 
cosmology on the parameters $\Omega_{\rm m}$ -- $\sigma_8$
[panel $a$)], 
$\Omega_{\rm m}$ -- $S_8=\sigma_8\left(\Omega_{\rm m}/0.3\right)^{0.5}$ [panel $b$)],
$\omega_{\rm m}$ -- $\sigma_{12}$ [panel $c$)], and 
$\omega_{\rm m}$ -- $S_{12}=\sigma_{12}\left(\omega_{\rm m}/0.14\right)^{0.4}$ [panel $d$)].}
\end{figure*}

\section{The normalization of the power spectrum}
Model predictions are often expressed in \hmpc units before AP distortions are 
taken into account. 
Using \hmpc or Mpc 
units yields identical cosmological constraints. 
However, \hmpc units obscure the response of $P(k)$
to changes in $h$.

Panel $a$) of Fig.~\ref{fig:normpk} shows the linear 
matter power spectra at $z=0$ of three $\Lambda$CDM models 
expressed in $h^{-1}{\rm Mpc}$ units, computed using CAMB \citep{Lewis2000}. 
These models have identical baryon, cold dark matter, and 
neutrino physical density parameters, $\omega_{\rm b}$, $\omega_{\rm c}$, and 
$\omega_{\nu}$, as well as scalar mode amplitude and spectral index, 
$A_{\rm s}$ and $n_{\rm s}$, and differ only in their values of $h$.
Panel $b$) of Fig.~\ref{fig:normpk} shows 
the same $P(k)$ in units of ${\rm Mpc}$, which 
have the same shape and differ only in their amplitude.
Expressing these power spectra in $h^{-1}{\rm Mpc}$ units 
obscures the fact that $h$ only 
affects the overall clustering amplitude.

\begin{figure}
\includegraphics[width=0.9\columnwidth]{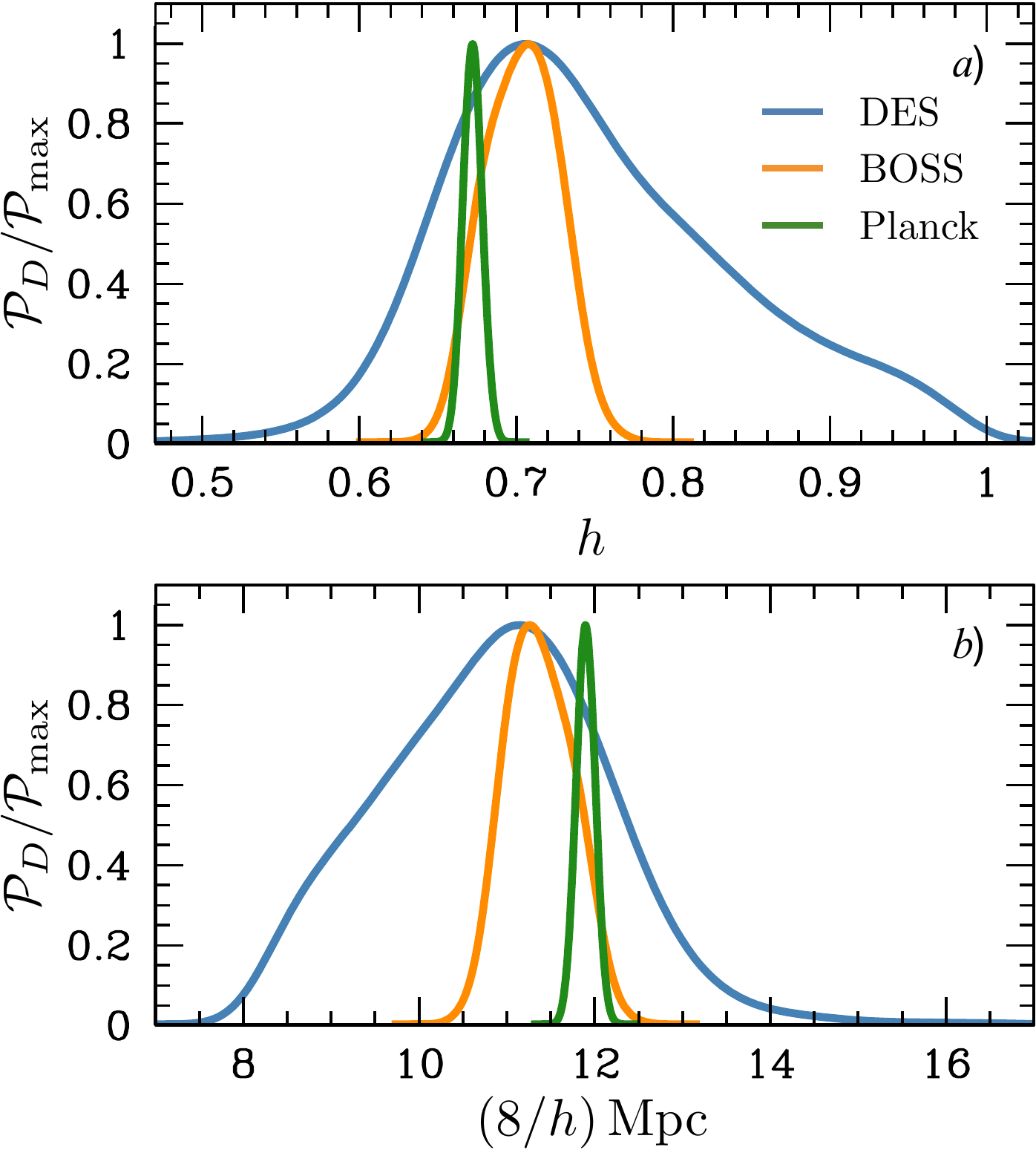}
\caption{\label{fig:posterior_h} 
Posterior distributions of the dimensionless Hubble parameter $h$ [panel $a$)] and 
the reference scale $(8/h)\,{\rm Mpc}$ where $\sigma_8$ is measured [panel $b$)]
recovered from DES, BOSS and Planck under the assumption of a $\Lambda$CDM 
universe.}
\end{figure}

For a $\Lambda$CDM universe, the amplitude of $P(k)$
is controlled by both $h$ and $A_{\rm s}$.
The joint effect of these parameters is usually described in terms 
of  $\sigma_8$. 
When $h$ varies, $\sigma_8$  
changes due to two effects:
 \begin{itemize}
 \item[\textit{i})] the change in the amplitude of $P(k)$ itself, and
 \item[\textit{ii})] the change in the reference scale $R=8\,h^{-1}\,{\rm Mpc}$, which 
 corresponds to a different scale in ${\rm Mpc}$ for different values of $h$.
 \end{itemize}
Point \textit{ii} implies that $\sigma_8$ does not capture the impact 
of $h$ on the amplitude of $P(k)$.
For different values of $h$, $\sigma_8$ characterizes
the amplitude of density fluctuations on different scales. 
Normalizing the power spectra of Fig.~\ref{fig:normpk} to the same value of $\sigma_8$
increases their amplitude mismatch. 

A better choice to describe the degenerate effect of $h$ and $A_{\rm s}$
is to normalize $P(k)$ using a reference scale in ${\rm Mpc}$. 
We propose to use  $\sigma_{12}$,
defined as the rms linear theory variance at $R= 12\,{\rm Mpc}$.
For models with $h\simeq 0.67$ as suggested by current CMB data, 
$8\,h^{-1}\,{\rm Mpc}\simeq 12\,{\rm Mpc}$, and $\sigma_{12}$ has a 
similar value to $\sigma_8$.
However, these parameters differ for other values of $h$. 
Panel $c$) of Fig.~\ref{fig:normpk} shows 
$P(k)$
for the same models of panel $b$) with their values of $A_{\rm s}$ 
modified to produce the same value of $\sigma_{12}$. These power spectra 
are identical, showing that the perfect degeneracy between $h$ and 
$A_{\rm s}$ is better described in terms of $\sigma_{12}$ 
than the standard $\sigma_8$. 

Panel $d$) of  Fig.~\ref{fig:normpk} shows the
nonlinear $P(k)$ of the same models as panel $c$), computed using HALOFIT
 \cite{Smith2008}.
The observed agreement, 
with differences of only a few percent at high $k$, 
shows that $\sigma_{12}$ is a more adequate parameter to 
characterize the nonlinear $P(k)$ 
than $\sigma_8$.

\begin{figure*}
\includegraphics[width=0.95\textwidth]{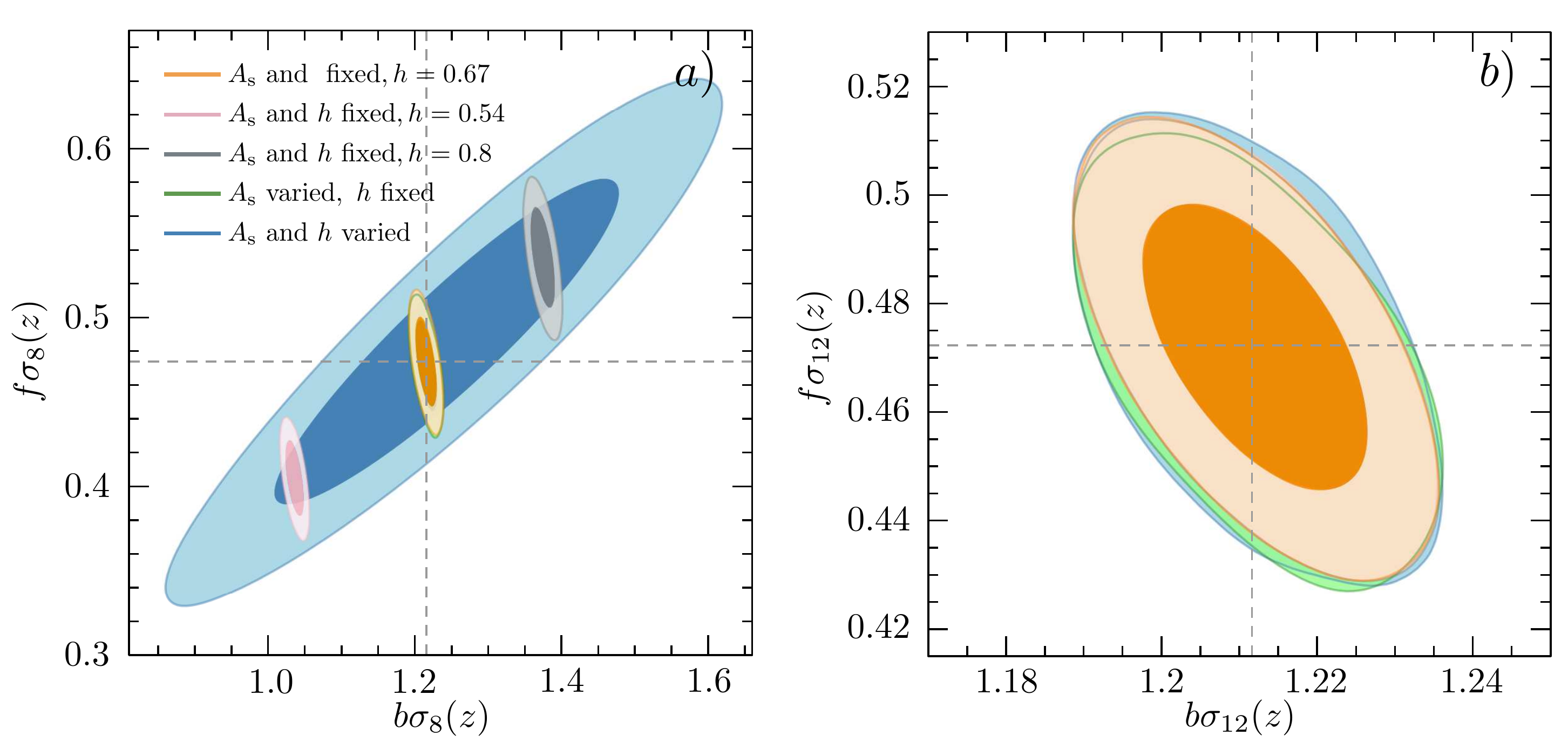}
\caption{\label{fig:rsd} 
Panel $a$): constraints on $b\sigma_8(z)$ and $f\sigma_8(z)$ derived from synthetic
Legendre multipoles $P_{\ell=0,2,4}(k)$. 
The contours correspond to the cases in which $A_{\rm s}$ and $h$ are fixed ($h=0.67$ orange, 
$h=0.54$ pink and $h=0.8$ gray), 
when $A_{\rm s}$ is varied  and $h$ is fixed (green), and when both  are varied (blue). Panel $b$): 
same constraints as panel $a$) but expressed in terms of $b\sigma_{12}(z)$ and $f\sigma_{12}(z)$.}
\end{figure*}

\section{Revising the $\sigma_8$ tension}
The value of $\sigma_8$ preferred by Planck CMB data \cite{Planck2018} under the 
assumption of a $\Lambda$CDM universe is higher than the estimates derived 
from all recent weak lensing (WL) datasets 
\cite{Heymans2012, Hildebrandt2017, Troxel2018, Hikage2018}
and the clustering measurements from BOSS \cite{Troster2020,Damico2019,Colas2019,Ivanov2019}. 
These discrepancies, dubbed the $\sigma_8$ tension, are 
illustrated in panel $a$) of Fig.~\ref{fig:consistency}, which shows the constraints on 
$\Omega_{\rm m}$ and $\sigma_8$ recovered from Planck \cite{Planck2018}, 
the auto- and cross-correlations between the  cosmic shear and galaxy positions 
from the Dark Energy Survey (DES) \cite{Abbott2018}, 
and clustering measurements from BOSS \cite{Sanchez2017,Troster2020}. 
These results assume  a $\Lambda$CDM cosmology with the same wide uniform
priors as in \cite{Troster2020}.
Panel $b$) of Fig.~\ref{fig:consistency} shows these constraints expressed in terms of 
$S_8 = \sigma_8\left(\Omega_{\rm m}/0.3\right)^{0.5}$. 
For the values of $\Omega_{\rm m}$ preferred by Planck, the low-redshift 
data prefer lower values of $S_8$ than the CMB. 

A drawback of using 
$\sigma_8$ to characterize the amplitude of $P(k)$
is that the reference scale $R=8\,h^{-1}{\rm Mpc}$ depends on $h$.  
Panel $a$) of Fig.~\ref{fig:posterior_h} shows the posterior distribution 
distribution on $h$, $\mathcal{P}(h)$, inferred from DES, Planck and BOSS.
Although they are consistent, DES gives a wider posterior than Planck or BOSS. 
The posterior $\mathcal{P}(h)$ impacts 
the constraints on $\sigma_8$, which are given by 
\begin{equation}
\sigma_8 = \int \sigma\left(R = \left(8/h\right)\,{\rm Mpc}\,|\,h\right) \mathcal{P}(h)\,{\rm d}h,
\end{equation}
that is, they represent the average of $\sigma(R)$ over the range of scales defined by the posterior 
distribution of $R=\left(8/h\right)\,{\rm Mpc}$, shown in panel $b$) of Fig.~\ref{fig:posterior_h}. 
Averaging $\sigma(R)$ over 
different scales will give different, not necessarily consistent, results. 

This issue can be avoided by using $\sigma_{12}$, which 
only depends on $h$ through its impact on the amplitude of $P(k)$.
Panel $c$) of Fig.~\ref{fig:consistency} shows the constraints in the 
$\omega_{\rm m}$ -- $\sigma_{12}$ plane 
recovered from the same data.
We use the physical density $\omega_{\rm m}$ instead of 
$\Omega_{\rm m}$ as the former is the most relevant quantity to characterize 
the shape of $P(k)$. When expressed in terms of $\sigma_{12}$, the 
constraints inferred from DES and Planck are in excellent agreement.
BOSS data prefer lower values of $\sigma_{12}$. 
Panel $d$) of Fig.~\ref{fig:consistency} shows these results in terms of 
the parameter
$S_{12} = \sigma_{12}\left(\omega_{\rm m}/0.14\right)^{0.4}$, 
which matches the degeneracy between 
$\omega_{\rm m}$ and $\sigma_{12}$ recovered from DES 
data. Planck and DES imply $S_{12} = 0.815\pm 0.013$ and  $S_{12}=0.798\pm 0.043$ 
respectively, while BOSS gives $S_{12}=0.716\pm 0.047$.
A detailed assessment of the consistency between Planck 
and low-redshift data is out of the scope of this work. 
However, such studies should characterize the amplitude of density 
fluctuations in terms of $\sigma_{12}$. 

\section{The growth rate of cosmic structures}
The analysis of redshift-space distortions (RSD) 
on clustering measurements 
is considered as one of the most robust 
probes of the growth-rate of structures \cite{Guzzo2008}. 
In linear perturbation theory, the relation between the two-dimensional 
galaxy power spectrum, $P_{\rm g}(k,\mu,z)$, and the real-space matter 
power spectrum can be written as \cite{Kaiser1987}
\begin{equation}
P_{\rm g}(k,\mu,z)= \left(b\sigma_8(z)+f\sigma_8(z)\mu^2\right)^2\frac{P(k,z)}{\sigma^2_8(z)}.
\label{eq:kaiser_fs8}
\end{equation}
where $\mu$ represents the cosine of the angle between ${\bf k}$ 
and the line-of-sight direction, $b(z)$ is the galaxy bias factor and 
$f(z)$ is the linear growth rate parameter. 
If $\sigma^2_8(z)$ described the amplitude of the power spectrum, the ratio 
$P(k,z)/\sigma^2_8(z)$ would
only depend on the parameters that control its shape. 
In this case, the anisotropies in $P_{\rm g}(k,\mu,z)$ 
would depend on the combination $f\sigma_8(z)$.
For this reason, 
the results of RSD analyses are usually expressed as measurements of $f\sigma_8(z)$.
However, this argument is flawed, as the ratio 
$P(k,z)/\sigma^2_8(z)$ depends on $h$. 
Instead, the ratio $P(k)/\sigma_{12}^2(z)$ is truly constant, independently of the 
values of $h$ or $\sigma_{12}$. Hence, the argument usually applied to justify the use 
of $f\sigma_8$ actually implies that $f\sigma_{12}$ 
is the most relevant quantity to describe RSD.

In most RSD studies, $f\sigma_8(z)$ is constrained together with 
the baryon acoustic oscillation (BAO) shift parameters, which describe 
the impact of AP distortions on the sound horizon scale,
while the cosmological parameters that determine the shape and amplitude 
of the matter $P(k)$, including $h$, are kept fixed. 
We can then expect to obtain different results
depending on the assumed value of $h$ or when this parameter is
marginalized over.
To illustrate this point, we 
used linear theory to compute the Legendre multipoles $P_{\ell=0,2,4}(k)$
of a galaxy sample roughly matching the volume, bias, and number density of the 
BOSS CMASS sample \cite{Reid2016} and used a Gaussian prediction
for their covariance matrix \cite{Grieb2016}. We
used these data to constrain $b\sigma_8(z)$, 
$f\sigma_8(z)$, and the BAO shift parameters.
Panel $a$) of Fig.~\ref{fig:rsd} shows the constraints in the 
$b\sigma_8(z)$ -- $f\sigma_8(z)$ plane obtained when 
both $A_{\rm s}$ and $h$ are kept fixed to their true values (orange), 
when $A_{\rm s}$ is varied while $h$ is kept fixed (green), and 
when both $A_{\rm s}$ and $h$ are varied (blue). 
The dashed lines indicate the true values of these parameters.
When only $A_{\rm s}$ is varied, the constraints follow the 
degeneracies defined by constant values of  
$b\sigma_8(z)$ and $f\sigma_8(z)$, leading to identical results to the 
ones obtained when it is fixed. 
However, when $h$ is also varied, the constraints deviate 
significantly from those of the standard case. The uncertainties 
on $f\sigma_8(z)$ derived under a fixed $h$ are significantly underestimated. 
Furthermore, the results obtained when fixing $h$ depend on the particular 
value adopted. This is illustrated by the pink and gray contours in 
Fig.~\ref{fig:rsd}, which show the results obtained assuming 
values of $h$ that differ by $\pm 20\%$ from the true value $h=0.67$.

Panel $b$) of Fig.~\ref{fig:rsd} shows the same constraints as in panel $a$) but expressed 
in terms of $b\sigma_{12}(z)$ and $f\sigma_{12}(z)$.  
The results 
are the same irrespective of whether $A_{\rm s}$ or $h$ are kept fixed or marginalized over.
This shows that $f\sigma_{12}(z)$ provides a more correct description of
the information retrieved from the standard RSD analyses. 

\section{Conclusions}
Although the use of \hmpc units has no impact on the information content of
cosmological data, they have generated misconceptions related to the normalization of the 
matter power spectrum in terms of $\sigma_8$.
This parameter does not correctly capture the impact of $h$ on the amplitude of $P(k)$, 
which is better described in terms of a reference scale in ${\rm Mpc}$.
A convenient choice is $12\,{\rm Mpc}$, which results in 
a mass variance $\sigma_{12}$ with a similar value to the standard $\sigma_8$ for 
$h\sim 0.67$. 

The amplitude of density fluctuations inferred from 
low- and high-redshift data should be characterized in terms of $\sigma_{12}$, eliminating
the dependency of the reference scale $R=8\,h^{-1}{\rm Mpc}$ on 
the constraints on $h$. 
The results of standard RSD analyses are more correctly described in terms of 
$f\sigma_{12}(z)$, which changes the cosmological implications of most available 
growth-rate measurements.
We propose to abandon the traditional \hmpc units in the analysis of new 
surveys \cite{Laureijs:2011gra, Levi:2013gra}, and to 
replace $\sigma_8$ by $\sigma_{12}$ to characterize 
the amplitude of density fluctuations. 

\emph{Acknowledgments.---}
A. G. S. would like to thank Daniel Farrow,
Daniel Gr\"un, Catherine Heymans,
Jiamin Hou,  Martha Lippich and Agne Semenaite
for their help and useful discussions. 
Some of the figures in this work were created with  \texttt{getdist}, making use of the 
\texttt{numpy} \citep{Oliphant2006} and 
\texttt{scipy} \citep{Jones2001} software packages. 
This reasearch was supported by the Excellence Cluster ORIGINS, 
which is funded by the Deutsche 
Forschungsgemeinschaft (DFG, German Research Foundation) under 
Germany's Excellence Strateg - EXC-2094 - 390783311.


%

\end{document}